\documentclass[twocolumn]{aastex63}
\usepackage{multirow}
\usepackage[T1]{fontenc}
\usepackage{mhchem}
\usepackage{url}

\newcommand{\Htwo}{\(\textrm{H}_{2}\)}
\newcommand{\CO}{\(^{12}\textrm{CO}\)}
\newcommand{\thCO}{\(^{13}\textrm{CO}\)}
\newcommand{\CetO}{\(\textrm{C}^{18}\textrm{O}\)}
\newcommand{\CsvO}{\(\textrm{C}^{17}\textrm{O}\)}
\newcommand{\HtwoO}{\(\textrm{H}_{2}\textrm{O}\)}
\newcommand{\HtwoetO}{\(\textrm{H}_{2}^{18}\textrm{O}\)}
\newcommand{\COtwo}{\(\textrm{CO}_{2}\)}
\newcommand{\Osix}{\(^{16}\textrm{O}\)}
\newcommand{\Oeight}{\(^{18}\textrm{O}\)}
\newcommand{\Msun}{\(\textrm{M}_{\odot}\)}

\newcommand{\Lya}{Ly-\(\alpha\)}
\newcommand{\etO}{\(^{18}\rm{O}\)}

\submitjournal{AJL}

\begin{document}
\title{Water UV-shielding in the terrestrial planet-forming zone: 
 Implications for Oxygen-18 Isotope Anomalies in \HtwoetO{} Infrared Emission and Meteorites}

\shorttitle{\HtwoO{}-to-\HtwoetO{} Ratio in the Inner Disk}


\author[0000-0002-0150-0125]{Jenny K. Calahan}
\affiliation{University of Michigan, 323 West Hall, 1085 South University Avenue, Ann Arbor, MI 48109, USA }

\author[0000-0003-4179-6394]{Edwin A. Bergin}
\affiliation{University of Michigan, 323 West Hall, 1085 South University Avenue, Ann Arbor, MI 48109, USA }

\author[0000-0003-4001-3589]{Arthur D. Bosman}
\affiliation{University of Michigan, 323 West Hall, 1085 South University Avenue, Ann Arbor, MI 48109, USA }

\begin{abstract}

An understanding of abundance and distribution of water vapor in the innermost region of protoplanetary disks is key to understanding the origin of habitable worlds and planetary systems. Past observations have shown \HtwoO{} to be abundant and a major carrier of elemental oxygen in disk surface layers that lie within the inner few au of the disk. The combination of high abundance and strong radiative transitions leads to emission lines that are optically thick across the infrared spectral range. Its rarer isotopologue \HtwoetO{} traces deeper into this layer and will trace the full content of the planet forming zone. In this work, we explore the relative distribution of H$_2^{16}$O and \HtwoetO{} within a model that includes water self-shielding from the destructive effects of ultraviolet radiation.
In this Letter we show that there is an enhancement in the relative \HtwoetO{} abundance high up in the warm molecular layer within 0.1-10~au due to self-shielding of CO, \CetO{}, and \HtwoO{}. Most transitions of \HtwoetO{} that can be observed with \textit{JWST} will partially emit from this layer, making it essential to take into account how \HtwoO{} self-shielding may effect the \HtwoO{} to \HtwoetO{} ratio. Additionally, this reservoir of \HtwoetO{}-enriched gas in combination with the vertical ``cold finger'' effect might provide a natural mechanism to account for oxygen isotopic anomalies found in meteoritic material in the solar system.
\end{abstract}


\section{Introduction} \label{sec:intro}

It is expected that the vast majority of solar-type stars host an exoplanet within 1~au \citep{Johnson10, Mulders18}. As a result, it can be assumed that the inner gaseous region of protoplanetary disks must be a hospitable environment for active planet formation.  Typical temperatures place nearly all primary volatile carriers of carbon, nitrogen, and oxygen into the gas within surface layers providing a rich environment for astronomical observations and constraint
 \citep{Pontoppidan14}. Volatiles (i.e. CO, \HtwoO, \COtwo) are potential starting points for the creation of complex molecules and the ice-coatings of pre-planetary pebbles \citep{Wang05,Gundlach15}.  These simple pre-cursors have been widely observed towards gas-rich disks within the 0.1-10~au region around young stars using the \textit{Spitzer} space telescope \citep[i.e.][]{Carr08,Pontoppidan10,Salyk2011}. Mapping the spatial extent and abundance of volatile molecules within the inner disk is essential knowledge used to piece together the connection between the chemical reservoir of a gas-rich disk and the resulting planets. 

The \textit{JWST} mission will provide improved spectral resolution access to the inner disk region \citep[JWST-MIRI:~$\lambda/\Delta \lambda = 2000-3000;$][]{Rieke15}, as compared to \textit{Spitzer}'s InfraRed Spectrograph  \citep[$\lambda/\Delta \lambda = 600;$][]{Houck04}. Nearly fifty disks will be targeted in the first set of \textit{JWST} observations and will be used to enhance the study of emissions from volatiles in the terrestrial planet-forming zone.
One molecule of high scientific impact is \HtwoO{}. Based on previous observations and theoretical work, it is expected that \HtwoO{} will be in high enough abundance such that its emission is optically thick \citep{Carr08}. The lesser abundant isotopologue \HtwoetO{} is also observable with JWST instruments, and, with reduced optical depth for its emission lines, will be used to infer the abundance and distribution of \HtwoO{} using the ratio between \Osix{} and \Oeight{}.

The most frequently used value for \Osix{}/\Oeight{} is 550. This value comes from \citet{Wilson99}, and is the average observed ratio for five local ISM sources. This ratio has been seen to vary across the Galaxy following

\begin{equation}
     (^{16}\textrm{O}/^{18}\textrm{O})= (58.8 \pm 11.8) \textrm{D}_{\textrm{GC}} + (37.1 \pm 82.6)
\end{equation}

\noindent where \(\textrm{D}_{\textrm{GC}}\) is the distance of a source from the Galactic center \citep{Wilson94}. Within the environment of a protoplanetary disk, however, there are chemical processes that can alter this ratio. 

One such process is molecular self-shielding.
Self-shielding is a process via which molecules, such as CO, N$_2$, and H$_2$, can protect themselves further from the source of radiation from the destructive effects of photodissociation from ultraviolet photons \citep{vandishoeck88}.  These molecules are dissociated via a line process, as opposed to molecules with photodissociation cross-sections that are continuous at UV wavelengths \citep{Heays17}.  Thus, molecules closer to the radiation source can absorb UV photons allowing for the lines to become optically thick and, hence, self-shield molecules downstream from the source of UV photons.
CO self-shielding can occur in gas-rich disks leading to a relative overabundance of $^{12}\textrm{C}^{16}\textrm{O}$ compared to $^{12}\textrm{C}^{18}\textrm{O}$ and $^{12}\textrm{C}^{17}\textrm{O}$ in the warm disk region \citep{Miotello14} because the absorbing lines of the lesser abundant isotopologues have reduced opacity. This process can help explain relatively high C\Osix{}/C\Oeight{} ratios that have been observed in a handful of protoplanetary disks \citep[i.e.][]{Brittain05,Smith09} and potentially might contribute to isotopic anomalies detected in asteroids and comets as compared to our Sun \citep{Lyons05, Nittler12, Altwegg20}. 

\citet{BethellBerginWater} demonstrated that, because of fast formation rates in hot ($>$400~K) gas, water can also self-shield.
Further, this shielding is unique in that it not only shields itself, but shields over a wide range of wavelength space, similar to ozone in the Earth's atmosphere. Thus, water might also operate as a shield for other molecules, a process we call water UV-shielding.
 The combination of CO self-shielding and \HtwoO{} UV-shielding could produce an environment where the \HtwoO{}, \HtwoetO{}, and CO are protected while \CetO{} continues to photodissociate. In this \(^{18}\rm{O}\)-rich environment, \HtwoetO{} will form and continue to be protected, raising the relative abundance of \HtwoetO{} with respect to \HtwoO{} in specific surface layers.  This could have a strong impact on both the CO/\CetO{} and \HtwoO{}/\HtwoetO{} values, altering them drastically from what is expected within the average ISM. These ratios will be commonly used in future \textit{JWST} observations due to the expected high optical depth of CO and \HtwoO{} within the inner disk. Additionally, the spatial extent of any \(^{18}\rm{O}\)-rich environment relative to where observed \HtwoetO{} and \HtwoO{} lines emit from will have a strong effect on the determination of \HtwoO{} abundance within protoplanetary disks.   Having accurate isotopic conversation factors could be critical as it is possible that the inner disk could betray evidence of volatile enhancements due to pebble drift \citep[][]{Ciesla06, Banzatti20}.
 
One final consideration to the chemical make-up of the inner disk is Lyman-$\alpha$ (Ly-\(\alpha\)) radiation. The Ly-\(\alpha\) transition often contains the vast majority of the energy within the FUV range if present in a stellar spectrum of T Tauri stars \citep[][]{Herczeg02, Schindhelm12}. It is readily observed towards young stars that are actively accreting and posses a gas-rich disk. The Ly-\(\alpha\) transition occurs at 1216~\AA{} and can photodissociate water \citep{vanDishoeck06}. Ly-\(\alpha\) is not often taken into account in radiation transfer codes, due to its added complexity of not only scattering off of dust but also atomic Hydrogen, causing the \Lya{} photons to scatter isotropically below the hot atomic layer of the disk \citep[][]{Bethell11}. To explore the strength that both water and CO self-shielding will have on the oxygen isotopic ratio, Ly-\(\alpha\) needs to be accounted for.

This paper is a companion to \citet{Bosman22a}, \citet{Bosman22b}, and Duval et al. (2022, in prep.) \citet{Bosman22a} sets up protoplanetary disk models focused on the innermost region which take into account water UV-shielding and efficient chemical heating processes. These two processes are required in order for a thermo-chemical calculation to reproduce water line emission and observed excitation temperatures from \textit{Spitzer}. Using the same modeling setup, \citet{Bosman22b} shows that CO\(_{2}\)-to-\HtwoO{} ratios can also be achieved, strengthened by additional depletion of excess oxygen beyond the water snowline. Duval et al. (in prep) will focus on the impact of water UV-shielding on the chemical composition of the terrestrial planet forming region. 

In this paper we will explore the abundances of both \HtwoO{} and \HtwoetO{} in the inner most few au of protoplanetary disks where the excitation conditions are favorable for detection of rotational and vibrational emissions at mid-infrared wavelengths.  This exploration will use the \textit{Spitzer} observational legacy as a constraint \citep{Pontoppidan10, Salyk2011}, but look forward to JWST observations which will be observing systems where both Ly-$\alpha$ radiation and water UV-shielding will be present. 


\section{Methods}\label{sec:method}

\subsection{Thermo-chemcial modeling: \textit{DALI}}

DALI (Dust And LInes) is a physical-chemical code that accounts for radiation transfer, thermal and chemical calculations throughout a disk \citep{Bruderer12, Bruderer13}. It contains an isotope chemical network \citep{Miotello14} and accounts for molecular self-shielding. For this project, \HtwoO{} UV-shielding has been additionally accounted for, see \citet{Bosman22a} for detail. We utilize a model derived from the AS 209 and corresponding stellar spectrum from \citet{Zhang21}, and explore the effect of \HtwoO{} UV-shielding and \Lya{} radiation in flat vs. thick model and two levels of dust settling. We include the \etO{} isotopologue of every oxygen carrying species in our chemical network.   

After an initial radiative transfer calculation, thermal balance, and chemical abundance distribution determination, we account for \Lya{} radiation using another code, detailed in the following section. The results from the separate \Lya{} radiative transfer calculation are combined with the initial disk radiation field as calculated by DALI.  We then calculate a new thermal balance and chemical environment, containing the final distribution of \HtwoO{} and \HtwoetO{}.

Excitation data for both ortho- and para-\HtwoetO{} was compiled into a file identical to the format from the Leiden LAMDA database \citep{Schoier05}.\footnote{\url{https://home.strw.leidenuniv.nl/~moldata/}} Line excitation data was obtained via the HITRAN database \citep{Gordon22}\footnote{\url{https://hitran.org/}} and collisional data from the \(\rm{H}_{2}^{16}\rm{O}\) LAMDA file \citep{Faure08}. This was used for ray-tracing calculations with DALI to determine which lines may be observable and from where in the disk they emit.  

\subsection{Inclusion of Lyman Alpha}

The inclusion of the Ly-\(\alpha\) is derived from a radiation transfer code described by \citet{Bethell11}. Briefly, this code calculates the transport of both FUV-continuum and Ly-\(\alpha\) photons. FUV photons are solely affected by the dust distribution, while \Lya{} is additionally affected by resonance line scattering. An initial distribution of H and \Htwo{} is derived using the thermo-chemical calculation from DALI, and \Lya{} propagates throughout the distribution of gas and dust, scattered by atomic H in the H-rich layer, and eventually scattered and absorbed by dust. 

After this calculation, the \Lya{}-affected radiation field is combined with the DALI radiation field. The effects of a separate stellar input spectrum are normalized out of the \Lya{} radiative transfer results, and a depletion or enhancement factor across wavelengths covered by the \Lya{} line and its line wings is calculated. The original DALI radiation field contains \Lya{} emission as it would exist if \Lya{}-photons scattered normally off of small dust grains only, identically to continuum-UV photons.



\section{Results}\label{sec:results}

\begin{figure*}
    \centering
    \includegraphics[scale=.5]{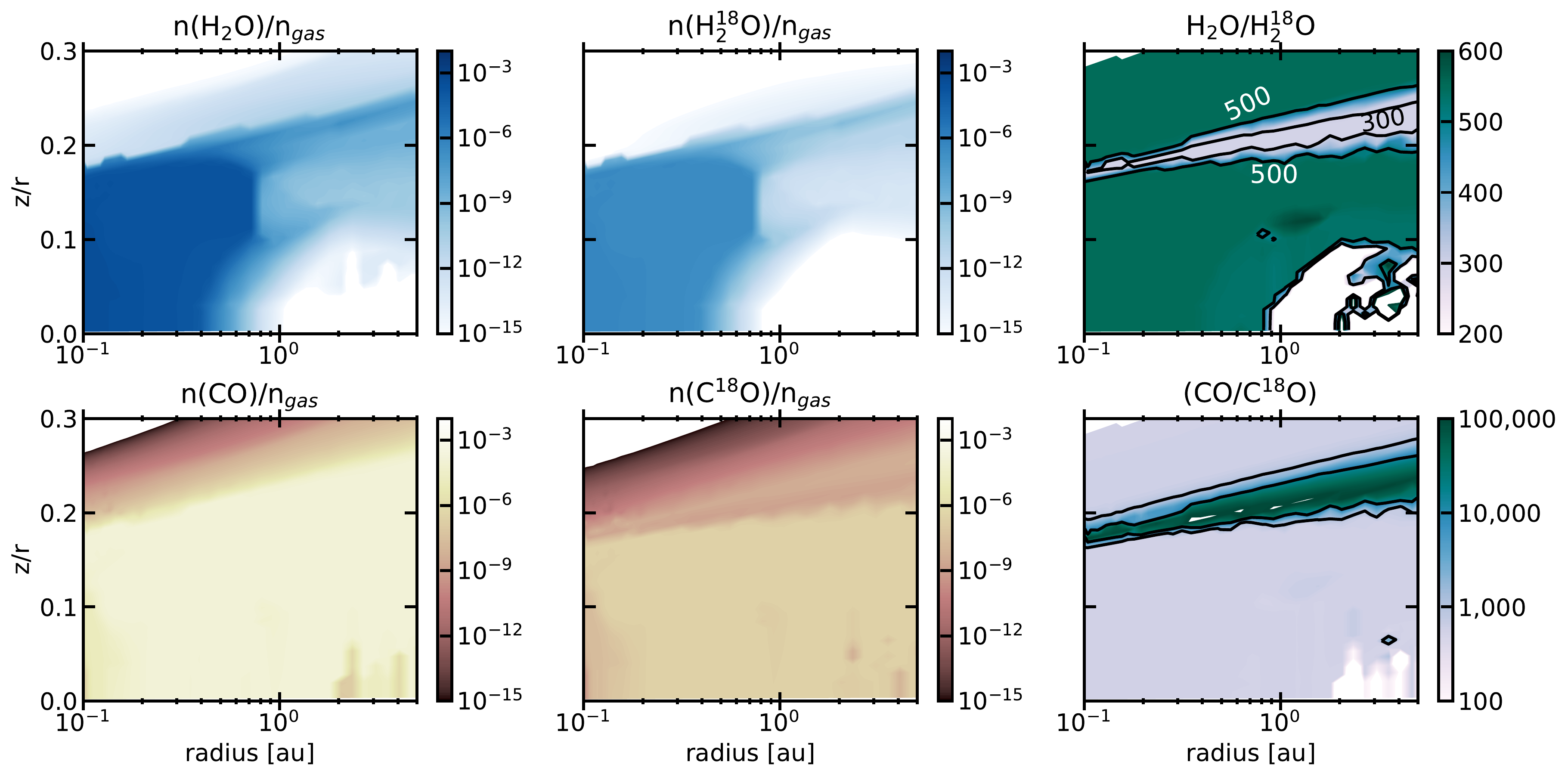}
    \caption{The \HtwoO{}, \HtwoetO{}, CO, and \CetO{} abundance in units of (mol/cm$^{-3}$)/n$_{gas}$ throughout the inner disk (left and center plots), and the ratio between \HtwoO{}, \HtwoetO{} (top right) and CO and \CetO{} (bottom right). At z/r$\approx{}$0.2 a unique region exits where \HtwoetO{} is enhanced and \CetO{} is greatly depleted. }
    \label{fig:o_o18}
\end{figure*}

\begin{figure}
    \centering
    \vspace{.1cm}
    \includegraphics[scale=.27]{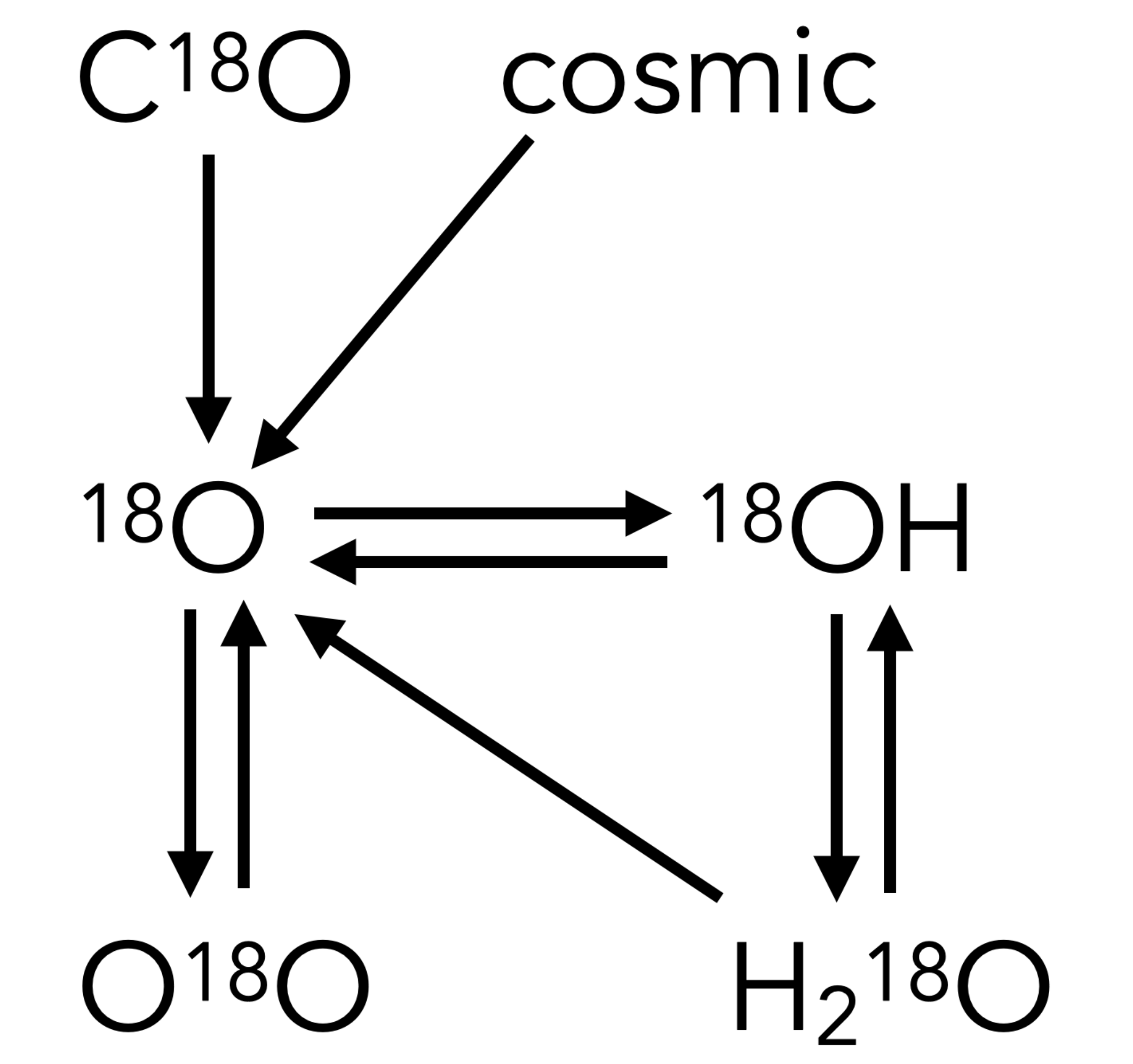}
    \caption{The evolution of \etO{} in equilibrium chemistry in the \HtwoetO{}-enhanced region. Most of the oxygen comes from the dark cloud material originating from molecules like H$_{2}$O, CO$_{2}$, organics and silicates. In this region photodissociation of \CetO{} provides and elevated abundance of \etO{} which reacts with H and \Htwo{} to quickly form $^{18}$OH and \HtwoetO{}. Because C$^{16}$O is shielded, very little extra $^{16}$O is included in this oxygen cycle, thus the relative increase in \HtwoetO{} production.}
    \label{fig:oxygen_cycle}
\end{figure}

\subsection{\HtwoetO{}-Enhanced Region}
We find an \HtwoetO{}-rich environment high up in the molecular region of the disk. Enhanced \HtwoetO{} exists in models that range in flaring angle and dust settling parameters. Four models were explored with two different values for scale height and dust settling, and these individual models are detailed in the Appendix of \citet{Bosman22a}. The following figures and discussion are based on the flattest and most highly settled model, which reproduce \HtwoO{} and CO$_{2}$ spectra simultaneously \citep[see][]{Bosman22b}. In this model, a region where \HtwoO{}/\HtwoetO{} dips to approximately 45\% of the initial ISM ratio exists at a z/r of 0.16-0.2 at r=0.1-1~AU, as seen in Figure \ref{fig:o_o18}. The upper boundary of this layer corresponds to where CO becomes optically thick enough to self-shield, producing an environment were \CetO{} continues to photodissociate while C$^{16}$O does not, releasing free \etO{}. Much of the free \etO{} finds its way into water molecules, enhancing the \HtwoetO{} abundance, thus providing a lower ratio between \HtwoO{} and \HtwoetO{}  (\(\approx{} 300\)). The most common \etO{} destruction mechanisms come in the form of interactions with \Htwo{} to form $^{18}$OH and an extra H atom. $^{18}$OH can then interact with an \Htwo{} atom to form \HtwoetO{}. The main cycle of \etO{} once the chemistry has reached equilibrium is shown in Figure \ref{fig:oxygen_cycle}. The ratio returns to an ISM level at z/r$\approx$0.16 when \CetO{} starts to be shielded, cutting of the supply of free \etO{}. Near the same z/r, within 1~au, \HtwoO{} UV-shielding aids in shielding \CetO{} to a point where its photodissociation rate is extremely low, many orders of magnitude lower than an environment where \HtwoO{} UV-shielding would be ignored.

Beyond \(\sim\)1~AU, the water abundance drops, yet a non-ISM ratio still exists. CO self-shielding continues to dominate, continuing to produce a \etO{} enhanced environment, thus a continued low ratio of \HtwoO{} to \HtwoetO{}, however there exists very little water vapor in the gas phase beyond r=1~au. The \HtwoetO{}-enhanced region exists between a top-down vertical CO column of $10^{17}$ and $10^{19}\rm{cm}^{-2}$, an \Htwo{} column of $10^{19}$ and $10^{22} \rm{cm}^{-2}$, and below an H column of $10^{21} \rm{cm}^{-2}$. This \HtwoetO{}-enhanced region co-exists with large temperature gradients. Within 1~au, the thermal range from the bottom of this layer to the top is 500-2,000~K. Beyond 1~au the gradient is larger, ranging from 100-4,000~K.


We find that most \Lya{} photons do not reach the \Htwo{}-dominated zone of the disk, including where the \HtwoetO{}-rich region exits. The \Lya{} emission feature is 25 times less bright in the radiation field of the disk at the location of the \HtwoetO{}-rich region than it would be if the transfer of \Lya{} photons was disregarded. Despite the decrease in overall UV flux, the \HtwoetO{}-rich region exits over the same spatial extent of the disk compared to a model that does not account for \Lya{} scattering off of hydrogen atoms. The limit at which scattered \Lya{} photons have a decreasing contribution to the radiation field corresponds to a z/r$\approx$0.25, or where the H-to-\Htwo{} ratio is $\approx10^{-5}$. While in this disk model \Lya{} did not reach into the molecular disk, \citet{Bethell11} found that the molecular disk can be enhanced with \Lya{} relative to FUV photons using the detailed disk physical models by \citet{DAlessio06}. In this study, we use the thermo-chemical calculation within DALI and a chemical network to determine the H-fraction throughout the disk, while \citet{Bethell11} used an analytical expression taking into account photodissociation and self-shielding processes while leaving out chemical destruction or creation pathways for \Htwo{} and H. This results in a H-to-\Htwo{} transition within the work that exists higher in the disk as compared to \citet{Bethell11}. Additionally, in \citet{Bethell11} the H-fraction drops to near zero while \Htwo{} is dominate while in DALI there exists a population of free H atoms in the \Htwo{}-dominated region of the disk due to destructive chemical reactions. Thus, in the DALI disk, there are scattering events deep in the disk where in \citet{Bethell11} there were none. The additional scattering events greatly increases the chance for \Lya{} photons to be absorbed by dust.



\subsection{H$_2^{18}$O Emission Spectrum}
\begin{figure*}
    \centering
    \includegraphics[scale=0.47]{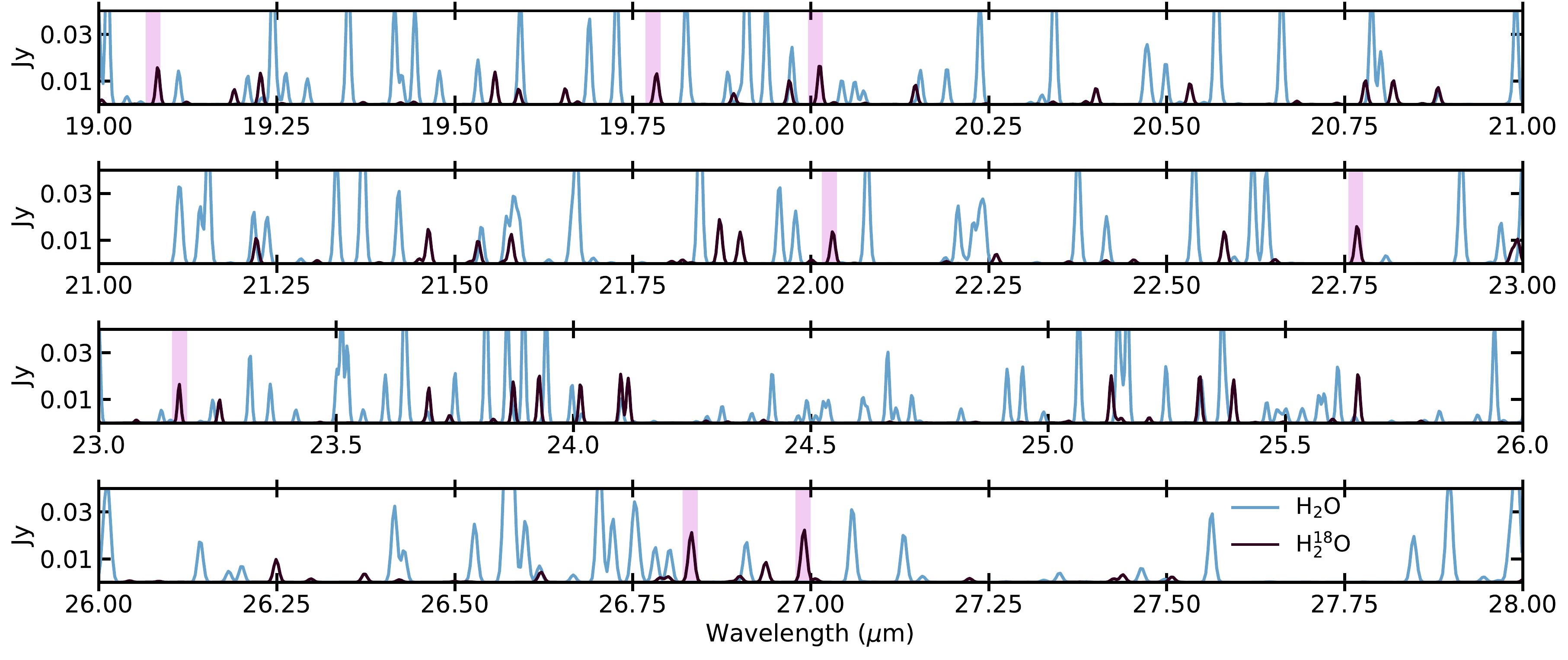}
    \caption{The \HtwoO{} and \HtwoetO{} spectra as predicted from an AS 209 model including water UV-shielding and \Lya{} contributions with 0.1 km s$^{-1}$ velocity bins. Blue lines correspond to \HtwoO{} and dark purple with \HtwoetO{}. \HtwoetO{} lines that are relatively bright and isolated from \HtwoO{} lines are highlighted. }
    \label{fig:water_lines}
\end{figure*}

The bulk of the water content exists within a radius of 1~au. 
We use DALI to calculate the flux contribution per cell for each emitted line. In order to determine which water lines will be of high interest, we run a quick non-LTE ray tracing program \citep[][Append. B]{Bosman17} that estimates the flux of water lines over the full extent of the MIRI wavelength range. We find that IR bright \HtwoO{} lines emit high up in the disk, proving to be optically thick across all IR wavelengths (see Bosman et al. 2022 in prep for detail). Isolated and bright \HtwoetO{} emission lines are rare, but a handful of lines will be accessible and distinguishable from \HtwoO{} with the MIRI instrument. We find eight distinct lines listed in Table \ref{tab:water_et_lines} and highlighted in Figure \ref{fig:water_lines}. These lines exist between 19-27~\micron{}; notably a region in MIRI with a lower sensitivity compared to shorter wavelength ranges. 

\begin{deluxetable}{ccccc}
\label{tab:water_et_lines}
\tablecolumns{3}
\tablewidth{0pt}
\tablecaption{Transitions of \HtwoetO{} Observable with JWST}
\tablehead{ &  & && \textbf{Weighted} \\
\textbf{Transition}&\textbf{Wavelength}  &\textbf{E$_{\textrm{up}}$} &\textbf{A}  & \textbf{$^{16}$O/$^{18}$O}\\
& \textbf{($\micron{}$)} & \textbf{(K)} & \textbf{(s$^{-1}$)}& \textbf{Average}}
\startdata
10$_{(3,8)}$ - 9$_{(0,9)}$ & 19.08 & 2072.3 &0.83& 520 \\
13$_{(5,8)}$ - 12$_{(4,9)}$ & 19.78 & 3772.5 &8.9& 500 \\
10$_{(4,7)}$ - 9$_{(1,8)}$ & 20.01 & 2265.3 &1.9& 510 \\
11$_{(4,7)}$ - 10$_{(3,8)}$ & 22.03 & 2725.3 &4.4& 490\\
9$_{(7,2)}$ - 9$_{(4,5)}$ & 22.77 & 2581.7 &0.031& 540\\
9$_{(9,0)}$ - 8$_{(8,1)}$ & 23.17 & 3165.9 &41.& 420\\
9$_{(6,3)}$ - 8$_{(5,4)}$ & 26.83 & 2329.5 &15.& 410\\
8$_{(7,2)}$ - 7$_{(6,1)}$ & 26.99 &  2265.6 &21.& 400\\
\enddata
\tablecomments{We quote each weighted average of the $^{16}$O-to-$^{18}$O ratio with two significant figures following the level of certainty around the ISM value (assumed to be 550).}
\end{deluxetable}

Each \HtwoetO{} line emits from distinct heights within the disk, and may partially emit from the \HtwoetO{}-enhanced region, thus when converting from \HtwoetO{}-emission to a \HtwoO{} abundance, a ratio below that of the ISM should be used. One of the brightest and most isolated lines is the \HtwoetO{} $8_{(7,2)}-7_{(6,1)}$ (27 \micron{}) transition. We find that this transition starts to become optically thick and emits primarily in the enriched \(^{18}\rm{O}\) region. We find this to be true for seven out of the eight identified lines. The \HtwoetO{} $9_{(7,2)}-9_{(4,5)}$ transition (22.77~\(\micron\)) appears to primarily emit below the \HtwoetO{}-enriched region, thus if using this line to determine a \HtwoO{} abundance, the conversion factor will be closer to ISM value of 550. 
\begin{figure*}
    \centering
    \includegraphics[scale=.5]{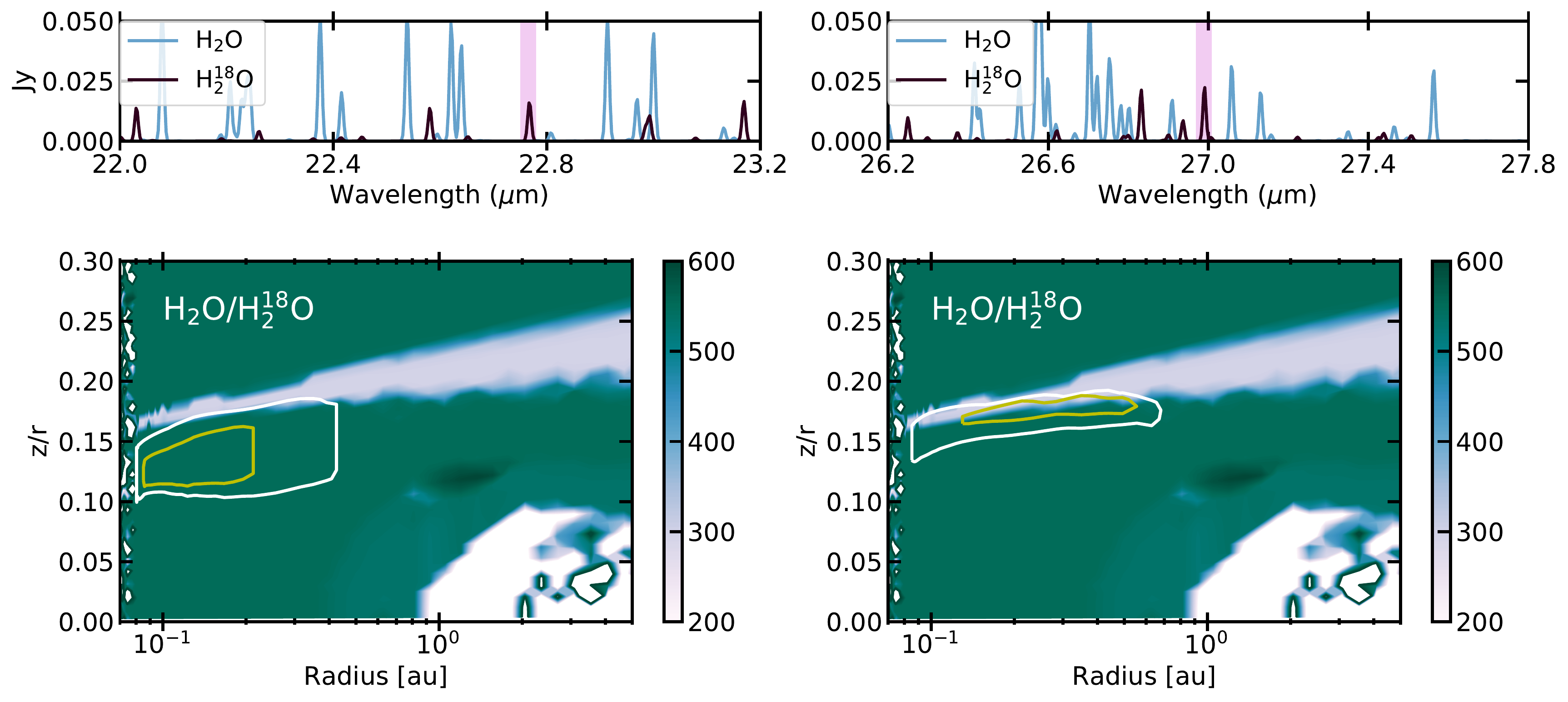}
    \caption{Top: The overlapping spectra from \HtwoO{} and \HtwoetO{} focusing on a narrow wavelength regions within the MIRI instrument on JWST, focused on the \HtwoetO{} 27 \micron{} line (left) and 22.77 \micron{} (right) lines. The bottom plots show the 50\% (yellow contour) and 90\% (white contour) contribution regions for each line plotted over the \HtwoO{}-to-\HtwoetO{} ratio.}
    \label{fig:27micron_example}
\end{figure*}


\section{Discussion \& Analysis}\label{sec:diss}

\subsection{Measuring \HtwoO{} abundance using \HtwoetO{} observations}\label{how_to_measure_water}

A weighted ratio between \HtwoO{} and \HtwoetO{} must be used in order to extrapolate to an \HtwoO{} abundance and distribution using \HtwoetO{} observations. We have identified eight transitions of \HtwoetO{} that are observable with JWST, and calculate the weighted ratio of \HtwoO{} to \HtwoetO{} that should be assumed based on our AS 209 model, and the vertical layers where we predict the emission arises. We compute weighted averages of \HtwoO{}/ \HtwoetO{} using the relative spatial contribution from where each line is emitting from, as calculated by DALI. 

Our first example comes from the  \HtwoetO{}~$8_{(7,2)}-7_{(6,1)}$ transition at 27 \micron{}. As seen in Figure \ref{fig:27micron_example}, much of the emission from this line originates from the \HtwoetO{} enhanced region. We use the following equation to calculated a weighted \HtwoO{}/\HtwoetO{} value:

\begin{equation}
    W = \frac{\sum_{i=1}^{n} w_{i} \chi_{i}}{\sum_{i=1}^{n} w_{i}}
\end{equation}

\noindent where \(n\) is the number of cells in our model, \(w_{i}\) is the relative emission contribution in a cell, and \(\chi_{i}\) is the \HtwoO{}/\HtwoetO{} value in a cell. Using this formalism, we find the average \HtwoO{}/\HtwoetO{} coming from this transition is \(\sim\)400. The \HtwoetO{} at 22.77 \micron{} appears to emit from a drastically different region in the disk, thus its average \HtwoO{}/\HtwoetO{} is 540, nearly identical within the margin of error to the typically used ISM estimate. We list the calculated weighted averaged determined for the eight identified bright and isolated \HtwoetO{} transitions in Table \ref{tab:water_et_lines}. There is a general trend that transitions from smaller wavelengths trend towards a weighted average of 550. This trend arises from the Einstein $A$-coefficient associated with each transition.  Transitions with lower $A$ values are more likely to emit from deeper within the disk at higher gas densities, thus farther away from the \HtwoetO{}-enhanced region.  To reduce the uncertainty in the \HtwoO{}/\HtwoetO{} ratio, transitions associated with low Einstein A-coefficients could be preferentially used. These differences are small and up to a factor of two.  However, chemical studies will be looking for enhanced abundances of water vapor in this region, perhaps supplied by pebble drift \citep{Banzatti20}.  To isolate these chemical changes we need to correct for effects that can be understood; water UV-shielding is one of these effects.


\subsection{Corresponding CO Infrared Observations}

In the \HtwoetO{}-enhanced region, \CetO{} abundance drops and the $^{12}$CO-to-\CetO{} ratio increases by upwards of four orders of magnitude. Measuring the $^{12}$CO-to-\CetO{} ratio using lines of CO that emit from this region could help confirm the presence of this \etO{}-rich region and even calibrate uncertainties in the true ratio between \HtwoO{}-to-\HtwoetO{}. The R and P branches for multiple vibrational transitions of CO and its isotopologues peak at $\sim$5 \micron{} and are observable from the ground using high spectral resolution instruments such as \textit{Keck} NIRSPEC and \textit{VLT}'s CRIRES. The \CetO{} column densities in the \HtwoetO{}-rich region range from 10$^{12}$-10$^{15}$ cm$^{-12}$ and are predicted to produce optically thin \CetO{} emission in the infrared. As summarized and expanded upon in \citet{Banzatti22}, over 100 T~Tauri and Herbig disks have been observed with high-spectral resolution in the wavelength range of CO's vibrational lines which emit from $\sim$1000~K, corresponding to the temperature of the \HtwoetO{}-enhanced region. In these studies only CO and \thCO{} observations were reported, making it currently impossible to use these observations to corroborate our predicted \etO{}-rich region within 1~au. The only work that set out to measure \CO{}/\CetO{} and \CO{}/\CsvO{} is found in \citet{Smith15}. This study observed nine YSO systems, three of which were optically thick disks. These three disks show tentative increase in the \CO{}/\CetO{} well above the ISM-measured value, however it was noted at significant systematic uncertainties are associated with these measurements. A future study dedicated to deep observations of multiple disk systems could aid in upcoming MIRI observations of these regions, and put constraints on the \etO{}-enhancement within the radial extent of planet-formation.

\subsection{Impact of a different $^{16}$O/\etO{} Ratio}

Mass-independent fractionation of oxygen isotopes, with enrichment of the heavy isotopes, has been isolated within meteoritic material \citep[][]{Clayton73, Theimens83}. The origin of these isotopic anomalies requires a specific mechanism that creates an \etO{} (and $^{17}$O)-rich environment as compared to the natal stellar nebula and envelope \citep[][]{Clayton93}. This fractionation must originate in the solar nebular disk or within the natal molecular cloud \citep[][]{Yurimoto04, Lyons05, Lee08}. CO self-shielding presents a mechanism to enrich gas in heavy isotopes, particularly on surfaces exposed to ultraviolet, as it selectively photodissociates \CetO{} and C$^{17}$O relative to CO producing gas enriched in \etO{} and $^{17}$O \citep{Lyons05, Miotello14}. In this work, we find that the addition of \HtwoO{} UV-shielding not only continues the \etO{} enrichment within the inner disk, but  enhances it compared to results solely based on CO self-shielding. We find a depletion of \CetO{} approaching a factor of 10$^{3}$ high up in the disk between 0.1-30 au at a $\sim$z/r=0.2-0.3. This is a reservoir rich with \etO{}, enhancing the \etO{} isotopologues for many different volatile species, including water vapor and ice.

Vertical mixing may act as a mechanism to deliver the \etO{}-enhanced molecules to the planet-forming midplane. The ``cold finger'' effect \citep{Stevenson88,Meijerink09} has been proposed a mechanism that will transport water and other organics efficiently from the atmosphere of the disk to the midplane. Due to vertical mixing, water from the gaseous atmosphere can transport down to the midplane where it locked onto grains as ice. The cold finger effect has been used to reconcile models in which high water abundances are predicted, yet corresponding sub-mm observations exhibit non-detections \citep{Du15, Salyk11,Carr18, Bosman21, Bosman22b}. Thus, while the \HtwoetO{}-rich regions resides high up in the disk atmosphere, vertical mixing can bring a portion of this reservoir down to the midplane and enrich meteoritic precursors.  



\citet{Lyons05} explored a solution in the outer tens of au that suggested that meteoritic \etO{} enrichment could be the result of CO self-shielding. In this model, oxygen atoms (enriched with heavy isotopes) produced via isotopic-selective CO self-shielding, are mixed downwards forming water ice via grain surface chemistry.  These ices will need to be transported to the inner few au, perhaps by drift \citep[][]{Ciesla06}.
However, current meteoritic evidence suggests the inner solar system is chemically separated from the outer ($\gtrsim$ 5 au) parts of the disk between 1~Myr and 3-4 Myr, potentially due to early formation of Jupiter's core \citep[][]{Kruijer17}. Our model shows that this vertical layer of enhanced \etO{} is placed directly into water in surface layers above the inner disk spatially closer to the region where meteoritic progenitors originate. 

\section{Conclusion}\label{sec:conclusion}

In this Letter, we implement \HtwoO{} UV-shielding and chemical heating in addition to CO self-shielding within a gas-rich disk environment. Focusing on the innermost region of the disk, a water vapor-rich reservoir exits within 1~au, and a \HtwoetO{} enhanced region exists where many water IR lines approach a \(\tau=1\). In this region, the \HtwoO{}-to-\HtwoetO{} ratio approaches 300, a significantly lower value than the assume ISM ratio of 550. We then seek use this context to provide insight into future JWST observations of \HtwoetO{} as it will act as a tracer for the abundance and distribution of gaseous \HtwoO{} in the main planet formation zone. 

\begin{enumerate}
    \item In our disk model, the \HtwoO{}-to-\HtwoetO{} ratio approaches 300 between a z/r$\approx$0.16-0.2 and inside 1~au due to CO self-shielding.
    
    
    \item We identify eight bright and isolated \HtwoetO{} lines that can be used to access the abundance and distribution of \HtwoO{}. Transitions associated with higher Einstein A-coefficients, such as \HtwoetO{}~$8_{(7,2)}-7_{(6,1)}$ at 27 \micron{} emit, primarily in the \HtwoetO{}-rich region, thus the ratio between \HtwoO{} and \HtwoetO{} are less than the local ISM value of 550. Seven out of the eight identified transitions emit predominately from the \HtwoetO{}-enhanced region, thus most observable \HtwoetO{} transitions with \textit{JWST} will have an additional factor of uncertainty when converting \HtwoetO{} abundance to \HtwoO{}.  
    
    \item In this model \Lya{} radiation does not penetrate down to the molecular-rich region of the disk in the case of our relatively thin and 0.0045 \Msun{} disk. 
    
    \item Vertical diffusive mixing, or the ``cold finger'' effect, could  transport a significant amount of \etO{}-rich molecules to the planet-forming midplane to be incorporated into pre-planetary materials that reside in the terrestrial planet-forming zone.
 
\end{enumerate}

\section{Acknowledgments}

\software{DALI, \citep{Bruderer12,Bruderer13} scipy \citep{scipy}, Matplotlib \citep{Matplotlib}, Astropy \citep{astropy:2013,astropy:2018}, NumPy \citep{harris2020array}, RADMC-3D \citep{Dullemond12}, radmc3dPy: \textsc{https://bitbucket.org/at\_juhasz/radmc3dpy/}, parallel \citep{parallel}}

J.K.C. acknowledges support from the National Science Foundation Graduate Research Fellowship under Grant No. DGE 1256260 and the National Aeronautics and Space Administration FINESST grant, under Grant no. 80NSSC19K1534. 

A.D.B. and E.A.B. acknowledge support from NSF Grant\#1907653 and NASA grant XRP 80NSSC20K0259


\bibliography{sample63}{}
\bibliographystyle{aasjournal}

\newpage

\end{document}